Tech Science Press

# A 360-Degree Panoramic Image Inpainting Network Using a Cube Map

## Seo Woo Han and Doug Young Suh*

Department of Electronic Engineering, Kyung Hee University, Youngin, 17104, South Korea
*Corresponding Author: Doug Young Suh. Email: suh@khu.ac.kr

**Abstract:** Inpainting has been continuously studied in the field of computer vision. As artificial intelligence technology developed, deep learning technology was introduced in inpainting research, helping to improve performance. Currently, the input target of an inpainting algorithm using deep learning has been studied from a single image to a video. However, deep learning-based inpainting technology for panoramic images has not been actively studied. We propose a 360-degree panoramic image inpainting method using generative adversarial networks (GANs). The proposed network inputs a 360-degree equirectangular format panoramic image converts it into a cube map format, which has relatively little distortion and uses it as a training network. Since the cube map format is used, the correlation of the six sides of the cube map should be considered. Therefore, all faces of the cube map are used as input for the whole discriminative network, and each face of the cube map is used as input for the slice discriminative network to determine the authenticity of the generated image. The proposed network performed qualitatively better than existing single-image inpainting algorithms and baseline algorithms.

## 1 Introduction

The consumption of images and videos increases exponentially as technology advances. People and devices not only consume images but actively generate images and videos. This trend has made an image and video editing and modification essential. An inpainting algorithm is a technique for restoring an image and removing unwanted objects based on information such as the texture or edges of an object [1]. Inpainting is used in many fields, such as image restoration [2,3], video transmission error repair [4], and image editing [5]. Inpainting has been a long-standing challenge in the field of computer vision [6]. Inpainting methods can be divided into non-learning-based methods and learning-based methods. Non-learning-based methods are divided into patch-based methods and diffusion-based methods. Patch-based methods [7,8] are used to fill a hole in an image by finding a similar pattern in an intact area within the image with which to fill the hole. Conversely, diffusion-based methods [9] fill a hole by successively filling in small portions from around the boundaries of the hole based on information gathered from the periphery of the hole. Elharrouss et al. [10] Non-learning-based methods do not require a dataset or training a network so that inpainted results can be obtained with less calculation. However, if the background doesn't feature a repeated pattern or the hole is exceptionally large, the inpainted results are not good [11]. To solve this problem, some researchers have studied learning-based methods. Methods using deep learning can be divided into those using convolution neural networks (CNNs) and those using generative adversarial networks (GANs). Recently, research on inpainting using GANs has been actively conducted because the features of GANs generate similar patterns based on the input image rather than simply using the information in the image, resulting in more plausible inpainted results. Generating datasets and training networks are time-consuming, but the inpainted results are more plausible compared to those of non-learning-based methods. However, the results of inpainting using deep learning are not good when filling exceptionally large holes or when the image features intricate patterns. Liu et al. [12] Research on image inpainting is continuing, and research on video inpainting is also actively underway. Video inpainting is difficult when applying an algorithm dedicated to image inpainting because video inpainting requires an accurate contextual understanding of frame and motion as well as the temporal smoothness of the output video [13]. Therefore, an inpainting algorithm using temporal and spatial information in video has been studied. Representative studies include a consistency-aware learning framework which simultaneously generates appearance and flow [14] and a method using high-quality optical flow [15].

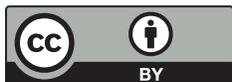

In this paper, we study a method of panoramic inpainting. The panoramic image, or panorama, is an image with a wide-angle of view. Panoramas are used for a variety of purposes, including landscape photography, group photography, and street views. Zhu et al. [16] Advances in camera technology have made it possible to shoot panoramic images, 360-degree panoramic images, and 360-degree videos without expensive professional equipment. 360-degree panoramas help create immersive content used to represent virtual reality (VR) and describe real space in a three-dimensional sense with a head-mounted display (HMD) used when viewing VR content. Very little research has been done on inpainting panoramic images. The inpainted result of a single-image inpainting algorithm on an equirectangular format panoramic image input is not good because the distortion of an equirectangular format panoramic image is not trained. Also, a memory shortage typically occurs during network training due to the very high resolution of equirectangular format panoramic images. To solve these two problems, we use a cube map format for the panoramic image inpainting instead of an equirectangular format.

The main contributions of this paper are as follows. First of all, a novel 360-degree panoramic image inpainting algorithm using deep learning is proposed. Instead of an equirectangular format, we use a cube map format with less distortion and propose a network structure which understands the correlation of the six sides of the cube map. Secondly, to train the cube map format panoramic image inpainting network, we use whole and slice discriminative networks trained to distinguish real images from inpainted results. The whole discriminative network looks at each entire cube map face to assess if it is correlative as a cube map image. The slice discriminative network looks at each face of the cube map to ensure local consistency. Finally, we validated the proposed network using 360-degree StreetView, the only publicly available 360–degree panoramic image dataset.

The paper is organized as follows: Section 2 briefly introduces the theoretical background. Section 3 explains the proposed model, and Section 4 describes the dataset used to train the proposed network. Section 5 describes the experiment and analysis with the proposed network, and Section 6 summarizes the results.

## 2 Related Works

This section briefly introduces a single image inpainting algorithm, a panoramic image inpainting algorithm, a conceptual description of a generative adversarial network, and research trends.

### 2.1 Single Image Inpainting

An inpainting algorithm can erase unwanted objects in an image or plausibly restore damaged or missing parts of an image. Inpainting technology is gradually diversifying from single images to videos. As mentioned in Section 1, inpainting is divided into non-learning-based methods and learning-based methods. Non-learning-based methods are effective in restoring texture but have difficulty in restoring overall shape. Yan et al. [17] For better performance than non-learning-based methods, an encoder-decoder network structure using CNNs has been proposed. Reference Zhu et al. [18] proposed a patch-based inpainting method for forensics images. U-nets Ronneberger et al. [19] and dense blocks were used to alleviate the gradient disappearance effect [20]. The latest trend in inpainting research is the use of GANs. A conceptual explanation of a GAN is given in Section 2.3. The use of only CNNs imposes many limitations because CNNs use only information from the input image. However, GANs can generate similar information based on the input image, so the inpainted result is more plausible than that of a method which only uses CNNs. Reference Liu et al. [21] proposed an inpainting method for faces using GANs. Since only GANs are used, the image resolution is low or tends to be unstable for training purposes. Therefore, many network structures using both CNNs and GANs have been proposed. In Nazeri et al. [22], a two-stage generator architecture was proposed to generate an image based on the edges around the hole in an image. After estimating the edge of the hole, the texture inside the edge is restored. The GAN-based inpainting method shows good performance but takes a long time to learn and has the disadvantage of requiring a high-performance training machine to calculate many parameters and perform many convolution operations.

### 2.2 Panoramic Image Inpainting

There are many ways to use inpainting algorithms on panoramic images. Just like an inpainting algorithm is used on a single image, it is used to erase unwanted objects in a panoramic image and reconstruct a damaged image. The study of panoramic inpainting has not progressed much compared to the study of single image inpainting. Zhu et al. [16] proposed a method for inpainting the lower part of a 360-degree panoramic image. This algorithm requires the projection map of the panoramic image. After the input image is projected onto a sphere, the lines and shapes are preserved and inpainted through matrix calculation. This algorithm inpaints only the lower part of the panoramic image, and the inpainted result is not good because it is not a learning-based method. Besides, it is limited in that it only works on images with simple patterns. Akimoto et al. [23] proposed an inpainting method using GANs using the symmetry of a 360-degree panoramic image. In this paper, there is no function to remove a specific object in an image. Only half of the buildings in a 360-degree panoramic street view image are used as input to the proposed network. This network restores a missing building by mirroring the building with the input image. After that, empty space is filled with plausible content. Uittenbogaard et al. [24] proposed the need for inpainting in a panorama to ensure privacy within a street view. This paper proposed a GAN-based inpainting algorithm using a multi-view of a 360-degree video with depth information which could detect and remove moving objects within the video. However, it has the limitation that it cannot be used on a single image. Also, to protect privacy, it provides results by blurring the detected object rather than erasing the object and filling its contents. Panoramic inpainting is also used in image extension technology that converts a single image into wide field-of-view images like a panoramic image. Extending images using existing inpainting algorithms leads to blurry results. To solve this problem, Teterwak et al. [25] proposed a panorama generation and image expansion technique by applying semantic conditioning to a GAN discriminative network.

### *2.3 Generative Adversarial Networks*

Generative adversarial networks Goodfellow et al. [26] have brought about tremendous advances in artificial intelligence. GANs are composed of two networks, as shown in Fig. 1: A generator, which creates new data instances, and a discriminator, which evaluates authenticity. The generator is called a generative network, and the discriminator is called a discriminative network. The generative network takes as its input a random vector $z$ to generate an image. At this time, the discriminative network receives the real image and the image created from the generative network as inputs to determine which image is real or fake. The goal of the adversarial network is to make the newly created data instance in the generative network and the real image indistinguishable to the discriminative network.

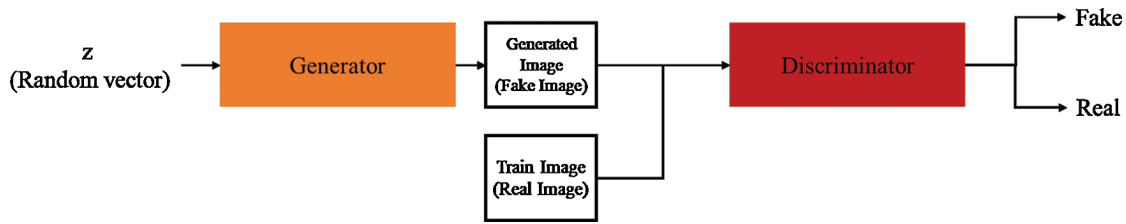

**Figure 1:** Generative adversarial networks architecture

For generative adversarial networks, the objective function satisfies Eq. (1). As in game theory, the two networks find a balance point with a single objective function.

$$\min_{G} \max_{D} V(G, D) = E_{x \sim P_{data}(x)}[\log D(x)] + E_{x \sim P_z(z)}[\log\{1 - D(G(z))\}] \tag{1}$$

Let the real data be $x$. The actual data distribution is $P_{data}(x)$ and the random vector distribution is $P_z(z)$. GANs learn to maximize the value function $V$ for $D$ and $G$ to minimize $\log\{1 - D(G(z))\}$. Discriminative networks are trained such that $D(G(z))$ is 0 and $D(x)$ is 1. The discriminative network trains to distinguish whether the input image is a generated image or a real image. The generative network trains the network so that the generated image is as similar as possible to the real image. Therefore, this structure is called a generative adversarial network because the generative and discriminative networks are trained as adversarial.

In conditional GANs, the input of the generative network is a random vector. Conditional generative adversarial networks (cGANs) Isol et al. [27] are complementary and modified structures which incorporate existing networks into images. cGANs train the mapping function from one image domain to another image domain and distinguish whether it is real or not through a discriminative network. The objective function of a cGAN satisfies Eq. (2). The first and second terms are the same as the existing GANs' objective function. $x$ and $y$ are paired. Let $x$ be the actual image and $y$ the label image. $z$ is a random vector used in the existing GANs.

$$\min_{G} \max_{D} V(G, D) = E_{x,y}[\log D(x, y)] + E_{x,z}[\log\{1 - D(x, G(x, z))\}] \tag{2}$$

Gulrajani et al. [28] confirmed that it is more effective to use the objective function of cGANs with traditional loss functions rather than simply using the objective function of cGANs. Therefore, the reconstruction loss function used in the CNN-based learning method was adopted. It was explained that the $L_1$ distance or the $L_2$ distance were used as a reconstruction loss function, and several tasks were tested, but the distance $L_1$ showed less blurry results than the distance $L_2$ and was used as the final objective function. The final objective function used in [28] satisfies Eq. (3)

$$\min_G \max_D V(G, D) = E_{x,y}[\log D(x, y)] + E_{x,z}[\log\{1 - D(x, G(x, z))\}] + \lambda E_{x,y,z}[||y - G(x, z)||_1] \qquad (3)$$

cGANs feature that input images and label images are input in pairs to the discriminative network. They also use the u-net structure as the generative network. Information loss occurs when using the encoder-decoder structure commonly used when dealing with images. The u-net is a structure in which an encoder-decoder structure adds a skip-connection which connects the corresponding encoder and decoder layers. Fig. 2 below shows the structure of cGANs and their difference from the original GANs. *x* is the real image, *y* is the label image paired with *x*, and $G(x)$ is the fake image created by the generative network.

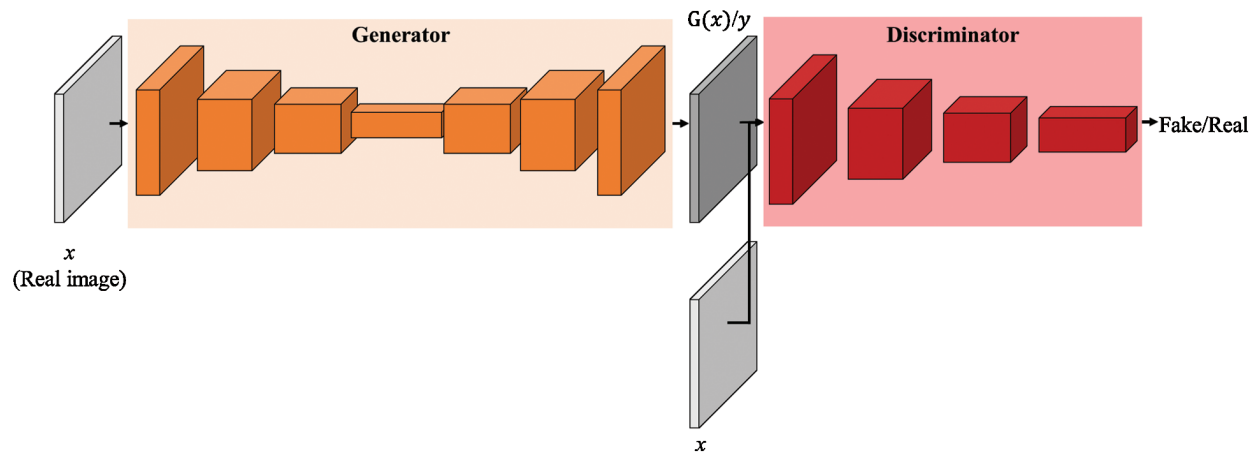

**Figure 2:** The conditional generative adversarial network architecture

## 3 Proposed Network

In this section, we describe the novel network structure and objective functions for the panoramic image inpainting. The input of the panoramic inpainting network proposed in this paper is an equirectangular format panoramic image and mask. When an equirectangular format panoramic image and mask are input, they are converted into a cube map format, then used as input to the generative network, which restores the damaged image.

In order to delicately restore the damaged part of each face of the cube map, it is input to a one-sided discriminative network. To train the correlation of the six sides, all six sides are input to the two discriminative networks at once. The output of the generative network is an inpainted image of the cube map format. While training on the panoramic image in the cube map format, we set the objective function suitable for this network using adversarial loss and reconstruction loss to obtain a plausibly inpainted result. The key parts of this paper are as follows. We used a cube map format with less distortion to inpaint the panoramic image. To train the texture of each image in the cube map, we designed a slice discriminative network which accepts as input one face of the cube map at a time. To train the correlation of the entire cube map, we designed a discriminative network which accepts as input all six faces of the cube map simultaneously. The proposed network structure is illustrated in Fig. 3.

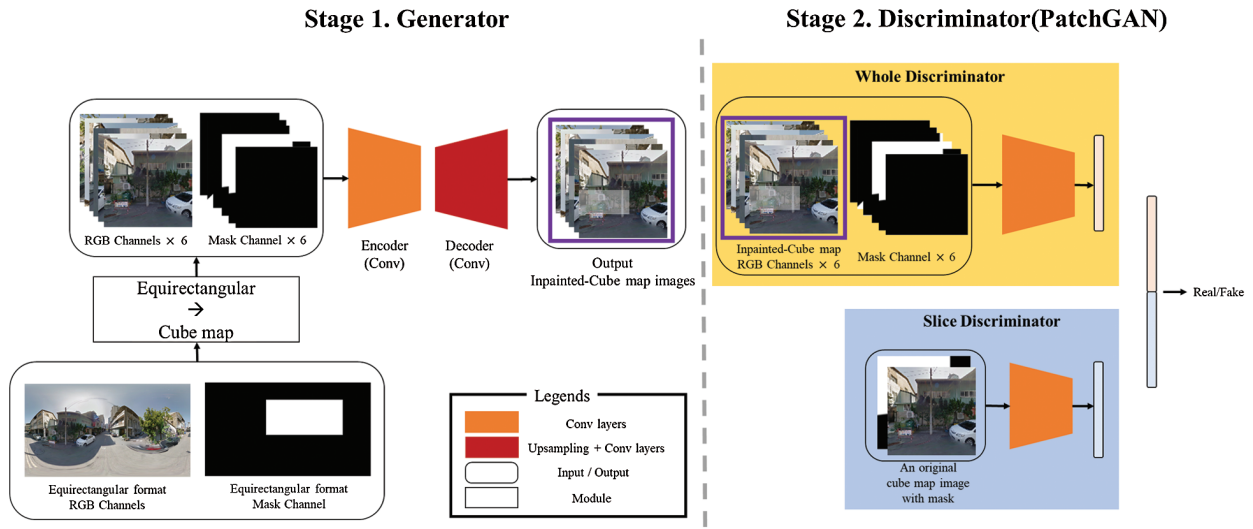

**Figure 3:** The proposed panoramic image inpainting network structure based on cGANs

### 3.1 Generative Network

The generative network is based on u-nets. A feature of a u-net is that it connects the encoder layer to the decoder layer, thus reducing the loss of image information. We modified the structure of the u-net to fit the cube map format image. We used LeakyReLU, ReLU, convolution (Conv.), transposed convolution (DeConv.), and batch normalization in the generative network. Tab. 1 shows the network structure of the generative network proposed in this paper.

**Table 1:** Generative network structure

| Generative network | | |
|---|---|---|
| Input: Damaged cube map images, masks | | |
| Output: Inpainted cube map images | | |
| Encoder | [Layer1] | Conv. Input channel = 4, output channel = 64, kernel size = 4, stride = 2, padding = 1; LeakyReLU; |
| | [Layer2] | Conv. Input channel = 64, output channel = 128, kernel size = 4, stride = 2, padding = 1; Batch norm; LeakyReLU; |
| | [Layer3] | Conv. Input channel = 128, output channel = 256, kernel size = 4, stride = 2, padding = 1; Batch norm; LeakyReLU; |
| | [Layer4] | Conv. Input channel = 256, output channel = 512, kernel size = 4, stride = 2, padding = 1; Batch norm; LeakyReLU; |
| | [Layer5] | Conv. Input channel = 512, output channel = 512, kernel size = 4, stride = 2, padding = 1; Batch norm; LeakyReLU; |
| | [Layer6] | Conv. Input channel = 512, output channel = 512, kernel size = 4, stride = 2, padding = 1; Batch norm; LeakyReLU; |
| | [Layer7] | Conv. Input channel = 512, output channel = 512, kernel size = 4, stride = 2, padding = 1; Batch norm; LeakyReLU; |

**Table 1 (continued).**

| | | |
|---|---|---|
| Decoder | [Layer8] | DeConv. Input channel = 512, output channel = 512, kernel size = 4, stride = 2, padding = 1; Batch norm; ReLU; Dropout = 0.5 |
| | | Concatenated Layer (Layer 8, Layer 6) |
| | [Layer9] | DeConv. Input channel = 1024, output channel = 512, kernel size = 4, stride = 2, padding = 1; Batch norm; ReLU; Dropout = 0.5 |
| | | Concatenated Layer (Layer 9, Layer 5) |
| | [Layer10] | DeConv. Input channel = 1024, output channel = 512, kernel size = 4, stride = 2, padding = 1; Batch norm; ReLU; |
| | | Concatenated Layer (Layer 10, Layer 4) |
| | [Layer11] | DeConv. Input channel = 1024, output channel = 256, kernel size = 4, stride = 2, padding = 1; Batch norm; ReLU; |
| | | Concatenated Layer (Layer 11, Layer 3) |
| | [Layer12] | DeConv. Input channel = 512, output channel = 128, kernel size = 4, stride = 2, padding = 1; Batch norm; ReLU; |
| | | Concatenated Layer (Layer 12, Layer 2) |
| | [Layer13] | DeConv. Input channel = 256, output channel = 64, kernel size = 4, stride = 2, padding = 1; Batch norm; ReLU; |
| | | Concatenated Layer (Layer 13, Layer 1) |
| | [Layer14] | DeConv. Input channel = 128, output channel = 3, kernel size = 4, stride = 2, padding = 1; |

### 3.2 Discriminative Network

The generative network is the same as that of [27], but the discriminative network is slightly different. The proposed network uses two discriminative networks. The whole discriminative network was made to be able to discriminate based on the correlation of the six sides of the cube map. The slice discriminative network was designed to determine whether inpainting was well done considering the texture of each side of the cube map. The channel size of the output layer is 1 for both the whole discriminative network and the slice discriminative network because the discriminative network must only discriminate whether its input is real or fake.

#### 3.2.1 Whole Discriminator

The whole discriminative network was used to train the correlation of each side of the cube map format panoramic image because when the inpainting is performed without considering the correlation of the six sides, a discontinuous image results when transformed into an equirectangular format. Tab. 2 shows the structure of the whole discriminative network used in this paper. Convolution, linear, LeakyReLU, and batch normalization are used for the whole discriminative network. The final output layer type of the entire discriminative network is (Batch number, 1).

#### 3.2.2 Slice Discriminator

The slice discriminative network was created to determine whether the input image for each side of the cube map format panoramic image is real or fake. The configuration of the slice discrimination network is illustrated in Fig. 4. The six cube map faces are sequentially entered into one slice discrimination network, and the outputs are combined into one. Therefore, the final output layer type of the slice discriminative network is (Batch number, 1).

**Table 2:** The whole discriminative network structure

| Whole discriminator network | |
|---|---|
| Input: Inpainted cube map images, masks | |
| Output: A transformed image of input (Batch number, 1) | |
| [Layer1] | Conv. Input channel = 24, output channel = 64, kernel size = 4, stride = 2, padding = 1; LeakyReLU; |
| [Layer2] | Conv. Input channel = 64, output channel = 128, kernel size = 4, stride = 2, padding = 1; Batch norm; LeakyReLU; |
| [Layer3] | Conv. Input channel = 128, output channel = 256, kernel size = 4, stride = 2, padding = 1; Batch norm; LeakyReLU; |
| [Layer4] | Conv. Input channel = 256, output channel = 512, kernel size = 4, stride = 2, padding = 1; Batch norm; LeakyReLU; |
| [Layer5] | Linear. Input channel = 32,768, output channel = 1 |

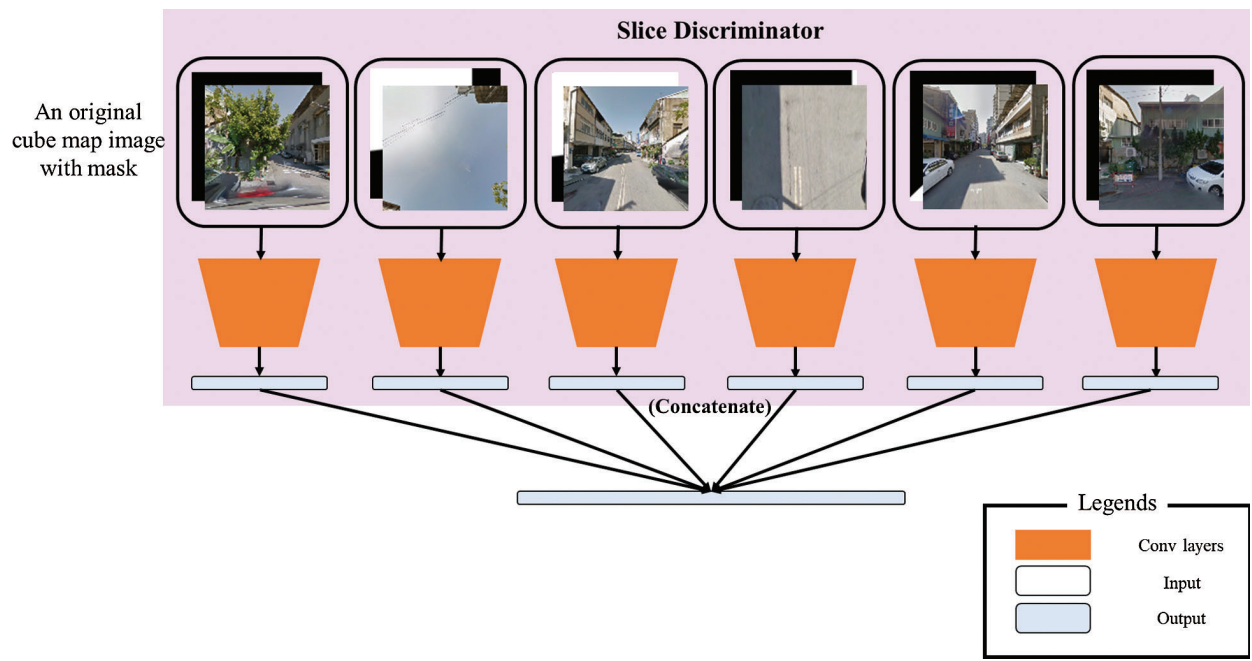

**Figure 4:** The slice discriminative network, which inputs the cube map one side at a time

Tab. 3 shows the structure of the slice discriminative network used in this paper. The slice discriminative network uses convolution, linear, LeakyReLU, and batch normalization.

**Table 3:** The slice discriminative network structure

| Slice discriminator network |
| --- |
| Input: An inpainted cube map image, masks |
| Output: Transformed images of input (Batch number, 1) |
| [Layer1] Conv. Input channel = 4, output channel = 64, kernel size = 4, stride = 2, padding = 1; LeakyReLU; |
| [Layer2] Conv. Input channel = 64, output channel = 128, kernel size = 4, stride = 2, padding = 1; Batch norm; LeakyReLU; |
| [Layer3] Conv. Input channel = 128, output channel = 256, kernel size = 4, stride = 2, padding = 1; Batch norm; LeakyReLU; |
| [Layer4] Conv. Input channel = 256, output channel = 512, kernel size = 4, stride = 2, padding = 1; Batch norm; LeakyReLU; |
| [Layer5] Linear. Input channel = 32,768, output channel = 1 |

### 3.3 Objective Function

The network proposed in this paper does not use the objective function of a cGAN described in Section 2. As mentioned in Gulrajani et al. [28], GANs are difficult to train. Methods to find ways to train GANs continuously are still being studied. To address the training difficulties of GANs, Gulrajani et al. [28] proposed the Wasserstein GAN gradient penalty (WGAN-GP). An existing Wasserstein GAN (WGAN) used the earth mover's distance (EMD) to calculate the distribution of generated data and real data. The objective function of a WGAN was created by applying the duality of Kantorovich-Rubinstein. The objective function of a WGAN in the generative network is Eq. (4), and the objective function of a WGAN of the discriminative network is to refer to Eq. (5).

$$\mathcal{L}_G^{WGAN} = -E_{\tilde{x} \sim \mathbb{P}_g}[D(\tilde{x})] \tag{4}$$

$$\mathcal{L}_D^{WGAN} = E_{x \sim \mathbb{P}_r}[D(x)] \tag{5}$$

Gulrajani et al. [28] developed a WGAN by adding a gradient penalty such as Eq. (6) to a WGAN. The points sampled in a straight line between the points sampled from the real data distribution $\mathbb{P}_r$ and the generated data distribution $\mathbb{P}_g$ is called $\hat{x}$.

$$\lambda_1 E_{\hat{x} \sim \mathbb{P}_{\hat{x}}}(||\nabla_{\hat{x}} D(\hat{x})||_2 - 1)^2 \tag{6}$$

In this paper, the objective function was defined by adopting ideas from Yu et al. [29], which was used by slightly modifying WGAN-GP. Since the image inpainting can be done by predicting the hole area in the image, the slope penalty is calculated using the product of the slope and the input mask **m**. It was modified and defined as in Eq. (7). ⊙ denotes the pixel product. If the mask value is 0, it is a damaged pixel; otherwise, it is 1.

$$\mathcal{L}_{gp} = E_{\hat{x} \sim \mathbb{P}_{\hat{x}}}(||\nabla_{\hat{x}} D(\hat{x}) \odot (1 - \mathbf{m})||_2 - 1)^2 \tag{7}$$

We used the weighted sum of the $l_1$ losses in the pixel direction and the adversarial losses in the WGAN. The $l_1$ loss function is Eq. (8) and the final objective function is Eq. (9).

$$\mathcal{L}_{l_1} = ||(1 - \mathbf{m}) \odot (G(\mathbf{x}, z) - \mathbf{x})||_1 \tag{8}$$

$$G^* = \arg \min_G \max_{D \in \mathcal{D}} (\lambda_1 \mathcal{L}_G^{WGAN} + \mathcal{L}_D^{WGAN} + \lambda_2 \mathcal{L}_{gp} + \lambda_3 \mathcal{L}_{l_1}) \tag{9}$$

In all experiments, $\lambda_1$ was set to 0.001, $\lambda_2$ was set to 10, and $\lambda_3$ was set to 1.2.

## 4 Dataset

This section shows an example of the dataset used in this paper. The proposed network trains network by using the image converted from the equirectangular format panoramic image to the cube map format panoramic image.

### 4.1 Image

In this paper, we used the street view equirectangular format panorama dataset provided in Chang et al. [30], which contains approximately 19,000 images. Because the images are street views, they can be divided into buildings and scenery. Let's call the building-rich images the building dataset and the tree-rich images the scenery dataset. There are 10,650 and 5080 images of buildings and scenery, respectively. In this paper, we confirmed the performance of the network with two building datasets and two scenery datasets.

As shown in Fig. 5a, when training with the equirectangular format panoramic dataset itself, the resolution of the panoramic images is high, resulting in a memory shortage, and the distortion of the equirectangular format panoramic images is challenging to train. To solve this, we lowered the resolution of the equirectangular format panoramic images and used images converted to a cube map format with relatively little distortion, as shown in Fig. 5b below.

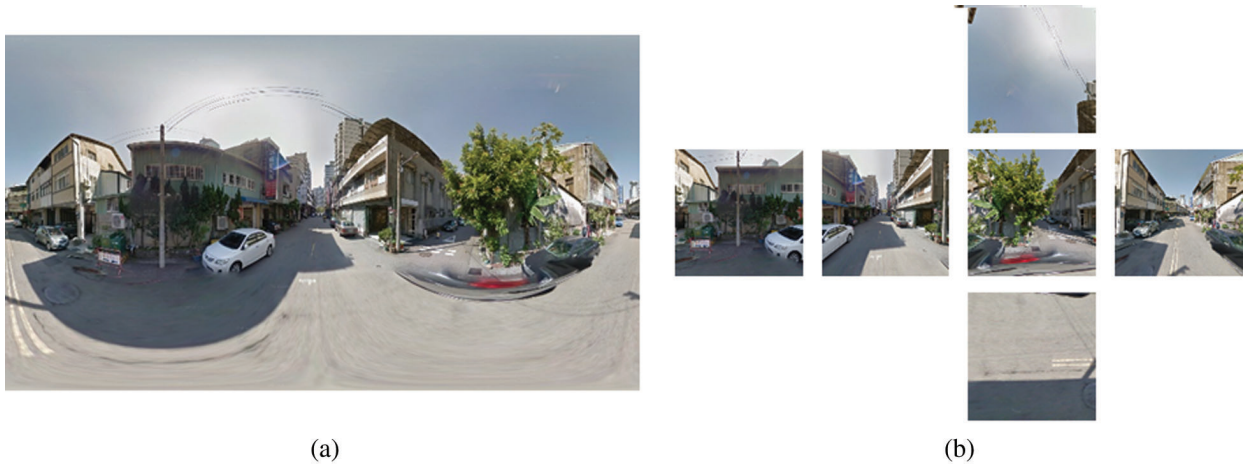

(a)                                                                                                         (b)

**Figure 5:** (a) Equirectangular format panoramic image and (b) Cube map format panoramic image

The panoramic image in the cube map format has six faces, as shown in Fig. 5b. Each side is referred to by a face name listed in Fig. 6. In this paper, the six faces are used in the following order: F, R, B, L, T, and D.

### 4.2 Mask

Most inpainting studies use two hole types to study inpainting: Rectangular masks in the form of Fig. 7a and free-form masks such as Fig. 7b are used to erase the shape of objects.

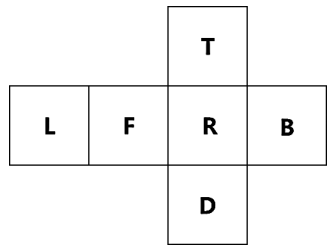

**Figure 6:** Cube map face names: L (Left), F (Front), R (Right), B (Back), T (Top), D (Down)

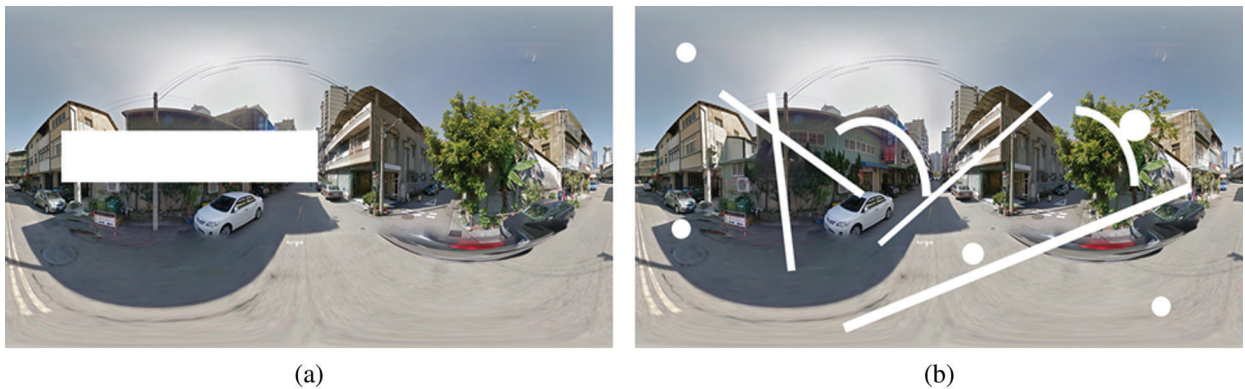

(a)                                                                                          (b)

**Figure 7:** (a) Rectangular mask and (b) Free-form mask

In this paper, our network used a rectangular mask because it has many applications, such as erasing or modifying objects and buildings, rather than delicately modifying the image, as do single-image inpainting algorithms. We made a rectangular hole before training a random number of images. Let the width and height of the panoramic image in equirectangular format be $w$ and $h$, respectively. Let the width and height of the rectangular hole be $R_w$ and $R_h$, respectively. The width and height of the hole used for training are expressed in Eq. (10).

$$\frac{w}{4} \leq R_w \leq \frac{w}{2} \; , \quad \frac{h}{4} \leq R_h \leq \frac{h}{2} \tag{10}$$

When the user edits the image, the constraints are set on the width and height of the rectangular hole, considering the size of the mask used. Also, in the cube map format, constraints were set to mask multiple faces of the cube map to train the correlation of the connected parts of each face. As shown in Fig. 8, the mask was also preprocessed to be converted from an equirectangular format to a cube map format. However, since the mask in the equirectangular format has no distortion, unlike the panoramic image in the equirectangular format, a straight line looks like a curve when converted to the cube map format.

## 5 Experiment and Analysis

In this section, we evaluate our method on one dataset: 360-degree StreetView. Since there is only one publicly available 360-degree street view image dataset, it was not possible to evaluate against various panoramic image datasets.

The system proposed in this paper uses a graphics processing unit (GPU) and is implemented using Pytorch. We measure the proposed panoramic inpainting system against a panoramic dataset [30]. Our model has full 7.3 M parameters and was trained on Pytorch v1.5 and CUDA v10.2. When learning the proposed network, the learning rate was set to 0.0004 and the batch size to 8. Even when validating the proposed network, the hole is Eq. (10). It is defined within the range according to Eq. (10), the equirectangular format panoramic image resolution is 512 × 256, and the cube map format panoramic image resolution is 256 × 256.

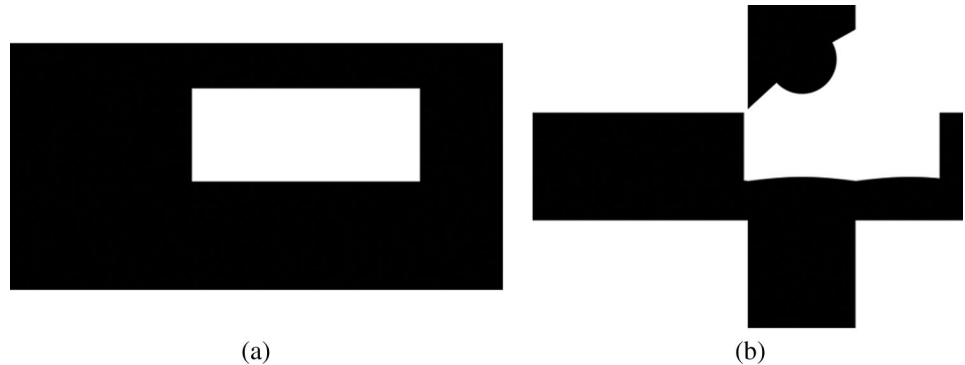

(a)                                                                     (b)

**Figure 8:** (a) Equirectangular format mask and (b) Cube map format mask

### 5.1 Qualitative Results

We compare our results with the state-of-the-art single image inpainting algorithm (GI) [29], a baseline using an equirectangular format panoramic image as input (OE), and a baseline using the cube map format panoramic image as input (OC). Our baseline models are comprised of cGANs. OE and OC used the cGAN network structure and objective function. OE is compared with our model to confirm that learning is difficult due to distortion when using an equirectangular format panoramic image. OC is compared with our model to check that the inpainted result is discontinuous when training the correlation of each side when using a cube map format panoramic image as input. GI, OE, and OC are implemented with Pytorch, and the hole size limitation is the same as in Eq. (9). The GI network was trained with a learning rate of 0.0001 and a batch size of 6, and the OE and OC networks were trained with a learning rate of 0.0002 and a batch size of 32. Fig. 9 shows the result of inpainting a 360-degree panoramic image using the proposed network. The first through fourth rows are the results of inpainting using the scenery dataset, and the fifth through eighth rows are results of inpainting using the building dataset. These are the plausible inpainted results when compared with the original images with masks and the inpainted images, which were the output of our proposed network.

Fig. 10 summarizes the qualitative results of the scenery dataset. The scenery dataset is relatively easy to inpaint because trees, roads, and sky are the main components of the images. Therefore, the inside of the mask is filled with very different objects than those in the original image. The palm tree trunks weren't well-erased in GI, OC, and OE, but they were in ours. Besides, in the case of OC, the correlation of each face of the cube map was not trained, so the boundary of each cube face is visible. GI and OE do not see the boundary like ours or OC because the equirectangular format panoramic image is used as input, but there is a distortion inherent in the equirectangular format, which confirms that the inpainted result is not natural.

Fig. 11 summarizes the qualitative results for the building dataset. The building dataset has more image components than the nature dataset, and the inpainting is difficult due to the presence of various buildings and roads in the images, and shadows caused by sunlight. Because GI uses a contextual attention mechanism, it is restored using similar colors and textures in the image. Therefore, it can be seen that similar results are used repeatedly, and the inside of the mask is restored. Ours and OC use the cube map format and convert it to an equirectangular format, so the cube map boundaries are visible and give implausible results. OC shows a blurry inpainted result, and OE is restored using plausible colors and textures, but when part of a building was expressed like a road, the result was implausible when the result is confirmed with the whole image.

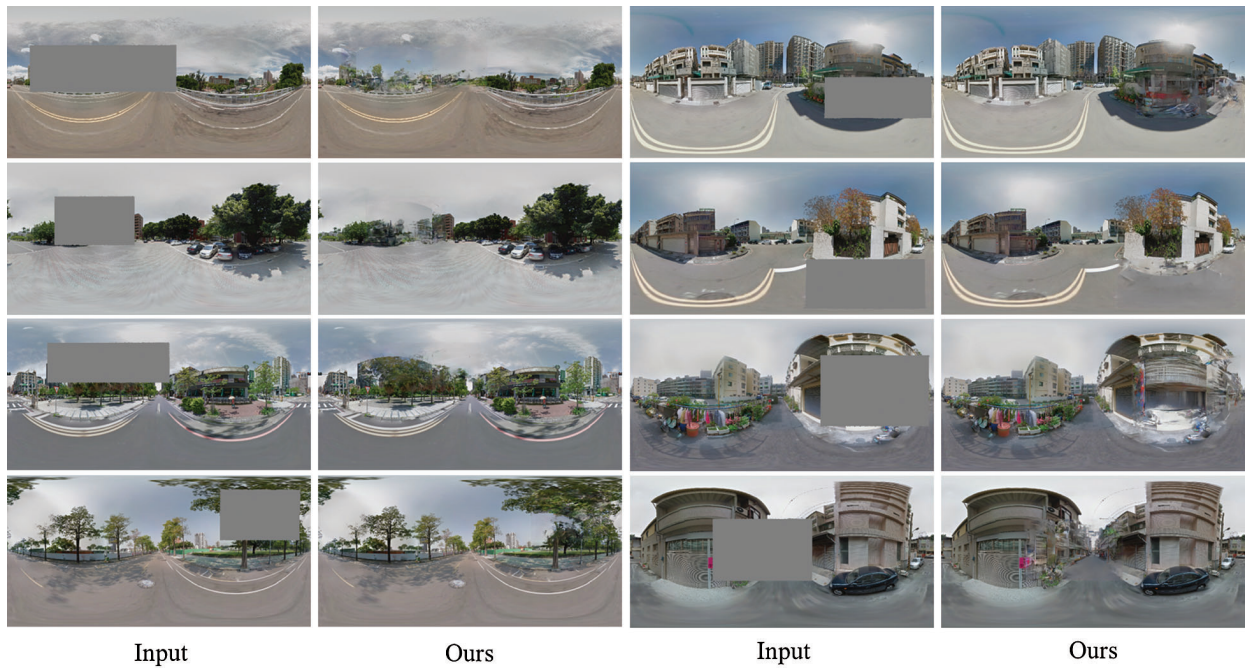

Input                      Ours                      Input                      Ours

**Figure 9:** Qualitative results using the scenery and buildings dataset

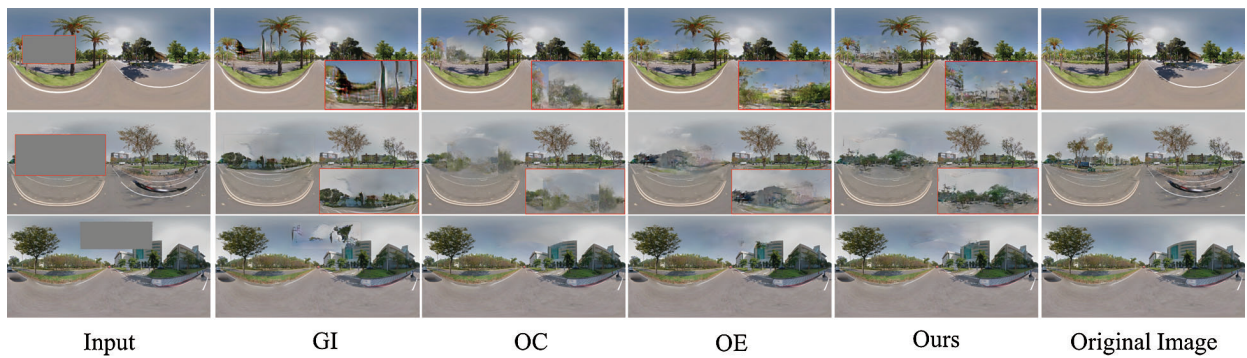

Input            GI             OC             OE             Ours          Original Image

**Figure 10:** Qualitative results using the scenery dataset

## 5.2 *Quantitative Results*

As mentioned in Yu et al. [29], image inpainting lacks a good quantitative evaluation scale. Structural similarity (SSIM), peak signal to noise ratio (PSNR), L1 distance and L2 distance values were compared for several algorithms and our proposed algorithm, regarding the evaluation metrics used in Yu et al. [29]. L1 distance and L2 distance find the pixel value difference from the original image. When GANs are used, the data distribution of the input image is learned to fill the empty hole in the image. The purpose of the inpainting study is to restore missing parts of an image plausibly. Therefore, L1 and L2 distances are relatively challenging to confirm network performance compared to SSIM and PSNR.

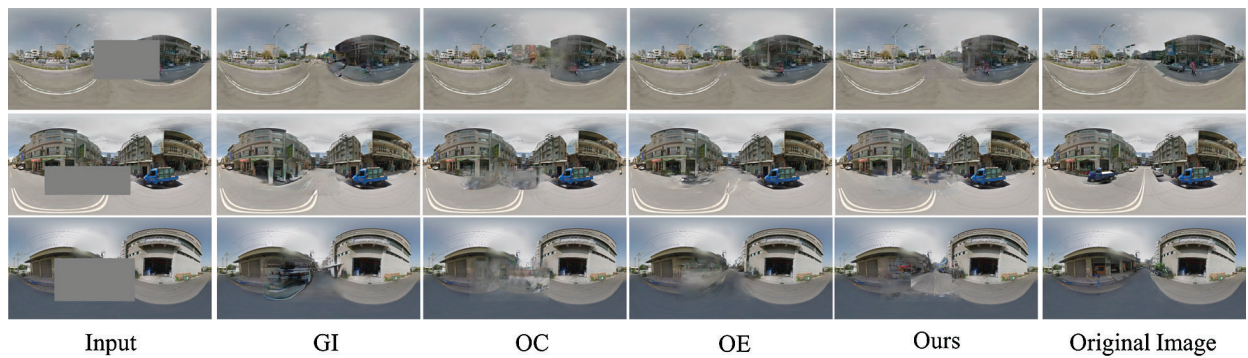

| Input | GI | OC | OE | Ours | Original Image |

**Figure 11:** Qualitative results using the building dataset

Tab. 4 shows that the OE model shows overall good performance. However, there is little difference between the metric value of the method we propose and the metric value of the OE model. Compared to a single image, a 360-degree panoramic image typically depicts various objects (e.g., trees, mountains, buildings, cars) in one image. Compared to the results of a single image inpainting algorithm, the panoramic image inpainting algorithm results may feature a variety of objects newly created by GANs and may differ significantly from the original image. For example, a road may be rendered in the space where a person or car had been deleted in a panoramic image. Therefore, the quantitative comparison of original and generated images is not a sufficient method of evaluating the models.

**Table 4:** Table of the quantitative results

|  | Scenery dataset (381 images) | | | | Buildings dataset (800 images) | | | |
|---|---|---|---|---|---|---|---|---|
|  | OC | OE | GI | Ours | OC | OE | GI | Ours |
| SSIM+ | 0.906 | **0.908** | 0.790 | 0.902 | 0.902 | **0.904** | 0.805 | 0.897 |
| PSNR+ | 37.7 | **37.8** | 32.4 | 37.7 | 37.7 | **37.7** | 32.7 | 37.4 |
| L1 distance− | 16.9 | **16.7** | 23.9 | 16.9 | **16.9** | 17.2 | 24.8 | 18.0 |
| L2 distance− | 11.5 | **11.2** | 21.2 | 11.4 | 12.3 | **11.7** | 26.2 | 12.3 |

## 6 Conclusion

We proposed a novel deep learning-based 360-degree panoramic image inpainting network. There is only one prior study of a deep learning-based 360-degree panoramic image inpainting method; Akimoto et al. [23] is an inpainting method using symmetric characteristics in an equirectangular format panoramic image, unlike a single image inpainting method. Therefore, only a limited number of 360-degree panoramic images can be inpainted with the network of Akimoto et al. [23], because the images must include a symmetrical building to be successfully inpainted. In contrast, the proposed network has the advantage of being able to inpaint like single-image inpainting methods by converting a panoramic image from an equirectangular format to a cube map format. However, since a plausible inpainted result is obtained only by training the correlation between the cube map format panoramic images, a panoramic image inpainting network is proposed comprised of a whole discriminative network and a slice discrimination network.

Training image inpainting networks using equirectangular format panoramic images is challenging because of distortion. To solve this problem, when using a cube map format panoramic image as input, it was confirmed that an additional algorithm or additional network layers were needed to train the correlation of each face of the cube map. Therefore, we obtained a plausible 360-degree panoramic image inpainted result by adding the whole discriminative network and the slice discriminative network to the baseline model. The whole discriminative network receives the six sides of the cube map as input simultaneously trains the correlation of the six sides of the cube map, and determines their authenticity. On the other hand, the six faces of the cube map are input one by one into the slice discriminative network to train the detailed texture of each face and to determine its authenticity.

The proposed network showed better qualitative and quantitative results than the single image inpainting algorithm. However, as mentioned in several image inpainting papers, there is no clear evaluation metric for comparing image inpainting performance. The L1 and L2 distances, which are traditionally used to evaluate image performance, are very inaccurate in evaluating the performance of the GANs because the original and generated images are compared. Although the proposed network did not show the best performance in quantitative results, it proved that it did not differ significantly from other networks. Besides, the proposed network produced the most plausible inpainted results through quantitative result images.

**Acknowledgement:** I would like to thank San Kim for his comprehensive advice and assistance in building and training networks. I would also like to thank my colleague Eun Young Cha for proofreading this article.

**Funding Statement:** This research was supported by Korea Electric Power Corporation (Grant No. R18XA02).

**Conflicts of Interest:** We declare that we have no conflicts of interest to report regarding the present study.

## References

[1]  Z. Yuan, H. Li, J. Liu and J. Luo, "Multiview scene image inpainting for intelligent vehicles based on conditional generative adversarial networks," *IEEE Transactions on Intelligent Vehicles*, vol. 2, no. 2, pp. 314–323, 2019.

[2]  X. Niu, B. Yan, W. Tan and J. Wang, "Effective image restoration for semantic segmentation," *Neurocomputing*, vol. 374, pp. 100–108, 2020.

[3]  D. Ren, W. Zuo, D. Zhang, L. Zhang and M. H. Yang, "Simultaneous fidelity and regularization learning for image restoration," *IEEE Transactions on Pattern Analysis and Machine Intelligence*, pp. 1–14, 2019.

[4]  S. Kamesh and K. R. Reddy, "Damaged video reconstruction using inpainting," in *Computer-Aided Developments: Electronics and Communication: Proc. First Annual Conf. on Computer-Aided Developments in Electronics and Communication*, 1st ed., vol. 1, Amaravati, India: Vellore Institute of Technology, pp. 201–207, 2019.

[5]  J. Youngjoo and J. Park, "Face editing generative adversarial network with user's sketch and color," in *Proc. IEEE Int. Conf. on Computer Vision*, Seoul, South Korea, pp. 1745–1753, 2019.

[6]  Y. Chen, Y. Tang, L. Zhou, Y. Zhou, J. Zhu *et al.*, "Image denoising with adaptive weighted graph filtering," *Computers, Materials & Continua*, vol. 62, no. 2, pp. 1219–1232, 2020.

[7]  R. Tijana and A. Pižurica, "Context-aware patch-based image inpainting using markov random field modeling," *IEEE Transactions on Image Processing*, vol. 24, no. 1, pp. 444–456, 2014.

[8]  Q. Guo, S. Gao, X. Zhang, Y. Yin and C. Zhang, "Patch-based image inpainting via two-stage low rank approximation," *IEEE Transactions on Visualization and Computer Graphics*, vol. 24, no. 6, pp. 2023–2036, 2017.

[9]  H. Li, W. Luo and J. Huang, "Localization of diffusion-based inpainting in digital images," *IEEE Transactions on Information Forensics and Security*, vol. 12, no. 12, pp. 3050–3064, 2017.

[10] O. Elharrouss, N. Almaadeed, S. AI-Maadeed and Y. Akbari, "Image inpainting: A review," *Neural Processing Letters*, vol. 51, pp. 1–22, 2019.

[11] S. Iizuka, S. S. Edgar and I. Hiroshi, "Globally and locally consistent image completion," *ACM Transactions on Graphics*, vol. 36, no. 4, pp. 1–14, 2017.

[12] G. Liu, F. A. Reda, K. J. Shih, T. C. Wang, A. Tao *et al.*, "Image inpainting for irregular holes using partial convolutions," in *Proc. European Conf. on Computer Vision*, Munich, Germany, pp. 85–100, 2018.

[13] C. Wang, H. Huang, X. Han and J. Wang, "Video inpainting by jointly learning temporal structure and spatial details," in *Proc. AAAI Conf. on Artificial Intelligence*, Honolulu, Hawaii, USA, vol. 33, pp. 5232–5239, 2019.

[14] H. Zhang, L. Mai, N. Xu, Z. Wang, H. Collomosse *et al.*, "An internal learning approach to video inpainting," in *Proc. IEEE International Conf. on Computer Vision*, Seoul, South Korea, pp. 2720–2729, 2019.

[15] R. Xu, X. Li, B. Zhou and C. C. Loy, "Deep flow-guided video inpainting," in *Proc. IEEE Conf. on Computer Vision and Pattern Recognition*, California, USA, pp. 3723–3732, 2019.

[16] Z. Zhu, R. R. Martin and S. M. Hu, "Panorama completion for street views," *Computational Visual Media*, vol. 1, no. 1, pp. 49–57, 2015.

[17] Z. Yan, X. Li, M. Li, W. Zhu and S. Shan, "Shift-net: image inpainting via deep feature rearrangement," in *Proc. European Conf. on Computer Vision*, Munich, Germany, pp. 1–17, 2018.

[18] X. Zhu, Y. Qian, X. Zhao, B. Sun and Y. Sun, "A deep learning approach to patch-based image inpainting forensics," *Signal Processing: Image Communication*, vol. 67, pp. 90–99, 2018.

[19] O. Ronneberger, P. Fischer and T. Brox, "U-net: Convolutional networks for biomedical image segmentation," in *Int. Conf. on Medical Image Computing and Computer-Assisted Intervention*, Munich, Germany, pp. 234–241, 2015.

[20] Y. Z. Su, T. J. Liu, K. H. Liu, H. H. Liu and S. C. Pei, "Image inpainting for random areas using dense context features," in *2019 IEEE Int. Conf. on Image Processing*, Taipei, Taiwan, pp. 4679–4683, 2019.

[21] H. Liu, G. Lu, X. Bi, J. Yan and W. Wang, "Image inpainting based on generative adversarial networks," in *2018 14th Int. Conf. on Natural Computation, Fuzzy Systems and Knowledge Discovery, IEEE*, Huangshan, China, pp. 373–378, 2018.

[22] K. Nazeri, E. Ng, T. Joseph, F. Qureshi and M. Ebrahimi, "Edgeconnect: Structure guided image inpainting using edge prediction," in *Proc. of the IEEE Int. Conf. on Computer Vision Workshops*, *Seoul, South Korea*, pp. 3265–3274, 2019.

[23] N. Akimoto, S. Kasai, M. Hayashi and Y. Aoki, "360-degree image completion by two-stage conditional gans," in *IEEE Int. Conf. on Image Processing*, Taipei, Taiwan, pp. 4704–4708, 2019.

[24] R. Uittenbogaard, C. Sebastian, J. Vijverberg, B. Boom, D. M. Gavrila et al., "Privacy protection in street-view panoramas using depth and multi-view imagery," in *Proc. of the IEEE Conf. on Computer Vision and Pattern Recognition*, California, USA, pp. 10581–10590, 2019.

[25] P. Teterwak, A. Sarna, D. Krishnan, A. Maschinot, D. Belanger *et al.*, "Boundless: Generative adversarial networks for image extension," in *Proc. of the IEEE Conf. on Computer Vision and Pattern Recognition*, Long Beach, California, USA, pp. 10521–10530, 2019.

[26] I. Goodfellow, J. P. Abadie, M. Mirza, B. Xu, D. W. Farley *et al.*, "Generative adversarial nets," in *Advances in Neural Information Processing Systems*, Montreal, Canada, pp. 2672–2680, 2014.

[27] P. Isola, J. Y. Zhu, T. Zhou and A. A. Efros, "Image-to-image translation with conditional adversarial networks," in *Proc. of the IEEE Conf. on Computer Vision and Pattern Recognition*, Honolulu, Hawaii, USA, pp. 1125–1134, 2017.

[28] I. Gulrajani, F. Ahmed, M. Arjovsky, V. Dumoulin and A. C. Courvile, "Improved training of wasserstein gans," in *Advances in Neural Information Processing Systems*, Long Beach, California, USA, pp. 5767–5777, 2017.

[29] J. Yu, Z. Lin, J. Yang, X. Shen and X. Lu *et al.*, "Generative image inpainting with contextual attention," in *Proc. of the IEEE Conf. on Computer Vision and Pattern Recognition*, Salt Lake City, Utah, USA, pp. 5505–5514, 2018.

[30] S. H. Chang, C. Y. Chiu, C. S. Chang, K. W. Chen, C. Y. Yao *et al.*, "Generating 360 outdoor panorama dataset with reliable sun position estimation," in *SIGGRAPH Asia Posters*, Vancouver, Canada, pp. 1–2, 2018.